\documentclass [preprint, prd, aps, nofootinbib]{revtex4}

\def\ve{\varepsilon}
\def\p{\partial}
\def\pslash{\p\raise.3ex \hbox{\kern-.5em /}}
\def\delslash{\nabla\raise.3ex \hbox{\kern-.7em /}}
\def\beq{\begin{equation}}
\def\eeq{\end{equation}}
\def\bea{\begin{eqnarray}}
\def\eea{\end{eqnarray}}

\begin{document}

\draft
\title{\large \bf Projective Geometry and $\cal PT$-Symmetric Dirac Hamiltonian}
\author{\bf Y. Jack Ng\footnote{E-mail: yjng@physics.unc.edu}
and H. van Dam}
\address{Institute of Field Physics, Department of Physics and
Astronomy,\\
University of North Carolina, Chapel Hill, NC 27599-3255\\}

\begin{abstract}

The $(3 + 1)$-dimensional (generalized) Dirac equation is shown to have the 
same form as the equation expressing the condition that a given point lies on 
a given line in 3-dimensional projective space.  The resulting Hamiltonian
with a $\gamma_5$ mass term is not Hermitian, but is invariant under the 
combined transformation of parity reflection $\cal P$ and time reversal
$\cal T$.  When the $\cal PT$ symmetry is unbroken, the energy spectrum of the free
spin-$\frac {1}{2}$ theory is real, with an appropriately shifted mass.

\bigskip

PACS numbers: 03.65.-w, 02.40.Dr, 03.65.Ta 

\end{abstract}

\maketitle

\bigskip

\section{Introduction and Summary}

Conventional quantum mechanics requires the Hamiltonian $H$ of any physical system  
be Hermitian (transpose $+$ complex conjugation)
so that the energy spectrum is real.  But as shown in the seminal paper by
Bender and Boettcher \cite{BB}, there is an alternative formulation of quantum
mechanics in which the requirement of Hermiticity is replaced by the condition of 
space-time $\cal PT$ reflection symmetry.  (For a recent review, see 
Ref.~\cite{review}.)
If $H$ has an unbroken $\cal PT$ 
symmetry, then the energy spectrum is real.  Examples of $\cal PT$-symmetric
non-Hermitian quantum-mechanical Hamiltonians include the class of Hamiltonians
with {\it complex} potentials: $ H=p^2+x^2(ix)^\epsilon$ with $\epsilon > 0$. 
Incredibly the energy levels of these Hamiltonians turn out to be real and 
positive. \cite{BB}  Now Hermiticity is an algebraic requirement whereas the 
condition of $\cal PT$ symmetry appears to be more geometric in nature.  
Thus one may wonder whether a purely geometric consideration can naturally 
lead to a Hamiltonian which is 
$\cal PT$-symmetric rather than its Hermitian counterpart.  
In this note we provide one such example.

In the next section, we ``derive'' the $(3 + 1)$-dimensional Dirac equation from a
consideration of the condition that a given point lies on 
a given line in 3-dimensional projective space.  By associating
the (homogeneous) coordinates of the point with the Dirac spinor components
$\psi({\bf x},t)$, and the coordinates of the line with the 
four-momentum and two real mass parameters $m_1$ and $m_2$ of the Dirac particle,
we are led to an equation taking on the form of a generalized Dirac equation
with Hamiltonian density  
\beq
{\cal H}({\bf x},t)=\bar\psi({\bf x},t)(-i\delslash+m_1+m_2\gamma_5)
\psi({\bf x},t)\quad(m_2~{\rm real}).
\label{e18}
\eeq
As noted in Ref. \cite{BJR}, the
Hamiltonian $H=\int d{\bf x}\,{\cal H}({\bf x},t)$
associated with the above ${\cal H}$ is not Hermitian but is invariant 
under combined ${\cal P}$ and ${\cal T}$ reflection.  For $\mu^2 \equiv 
m_1^2 - m_2^2 \geq 0$, it is equivalent to a Hermitian
Hamiltonian for the conventional free fermion field theory with mass 
$\mu$.  Studies of spin-${\frac {1}{2}}$
theories in the framework of projective geometry have been 
undertaken before.  See, e.g., Ref.~\cite{Dirac}. \footnote{These papers
are rather mathematical and technical.  The authors of the first two
papers discuss the Dirac equation  in terms of the Plucker-Klein
correspondence between lines of a three-dimensional projective space and
points of a quadric in a five-dimensional projective space.  The last
paper shows that the Dirac equation bears a certain relation to Kummer's
surface, viz., the structure of the Dirac ring of matrices is related to
that of Kummer's $16_6$ configuration.  All these authors, explicitly or
implicitly, put one of the two masses, viz., $m_2$ in (\ref{e18}), to be
zero by hand.  In this note, we ``derive" the generalized Dirac equation
from the projective geometrical approach in a relatively simple way and
point out that there is no need 
to put $m_2 = 0$ and perhaps it is even natural to keep {\it both} 
masses $m_1$ and $m_2$.}  But the idea that there may be 
a natural {\it connection} between the projective 
geometrical approach (perhaps also other geometrical approaches) 
and $\cal PT$-symmetric Hamiltonians as pointed out in this
note appears to be novel.


\bigskip

\section{Projective geometry and $\cal PT$-symmetric Dirac equation}

It is convenient to use homogeneous coordinates to express the geometry in a
projective space. \cite{PG}  A point ${\bf x}\equiv (x, y, z)$ in 
three-dimensional Euclidean space can be
expressed by the ratios of four coordinates $(x_1, x_2, x_3, x_4)$ which are called
the homogenous coordinates of that point.  
One possible definition of $(x_1, x_2, x_3, x_4)$,
in terms of ${\bf x}$ is $x_1 = \frac {x}{d}, x_2 = \frac {y}{d}, x_3 = \frac {z}{d},
x_4 = \frac {1}{d}$, with $d$ being the distance of the point from the origin. 
Obviously, for any constant $c$, $(cx_1, cx_2, cx_3, cx_4)$ and $(x_1, x_2, x_3, x_4)$ 
represent the same point ${\bf x}$.  

Consider the line through two points
$(a_1, a_2, a_3, a_4)$ and $(b_1, b_2, b_3, b_4)$.  For 
$(x_1, x_2, x_3, x_4)$ to lie on that line, the following determinant has to
vanish, for any $(r_1, r_2, r_3, r_4)$,
\begin{equation}
\left| \begin{array}{cccc}
         a_1 & b_1 & x_1 & r_1 \\
         a_2 & b_2 & x_2 & r_2 \\
         a_3 & b_3 & x_3 & r_3 \\
         a_4 & b_4 & x_4 & r_4
        \end{array}  \right|
      = 0.
\label{bigd}
\end{equation}
This gives, for any $r_1$,
\begin{equation}
\left| \begin{array}{ccc}
         a_2 & b_2 & x_2 \\
         a_3 & b_3 & x_3 \\
         a_4 & b_4 & x_4 
        \end{array}  \right|
      = 0.
\label{medd}
\end{equation}
With the aid of the Plucker coordinates of the line defined by
\begin{equation}
p_{ij}= - p_{ji} \equiv
\left| \begin{array}{cc}
        a_i & b_i \\
        a_j & b_j
       \end{array}  \right|,
\label{smalld}
\end{equation}
Eq. (\ref{medd}) can be written as
\begin{equation}
p_{34} x_2 - p_{24} x_3 + p_{23} x_4 = 0.
\label{r1}
\end{equation}
Similarly, for any $r_2, r_3, r_4$, the following equations respectively 
must hold
\bea
p_{41} x_3 - p_{31} x_4 + p_{34} x_1 = 0,\nonumber\\
p_{12} x_4 - p_{42} x_1 + p_{41} x_2 = 0,\nonumber\\
p_{23} x_1 - p_{13} x_2 + p_{12} x_3 = 0.
\label{r4}
\eea
Note that the Plucker line coordinates are not independent; the identical
relation that connects them can be found by expanding the determinant
\begin{equation}
\left| \begin{array}{cccc}
         a_1 & b_1 & a_1 & b_1 \\
         a_2 & b_2 & a_2 & b_2 \\
         a_3 & b_3 & a_3 & b_3 \\        
         a_4 & b_4 & a_4 & b_4
        \end{array}  \right|
      = 0,
\label{bigd1}
\end{equation}
from which
\begin{equation}
p_{12}p_{34} + p_{13}p_{42} + p_{14}p_{23} = 0.
\label{relation1}
\end{equation}

Next, we relabel the homogeneous coordinates $(x_1, x_2, x_3, x_4)$ as
the four Dirac spinor components $\psi$.  
Let us first use the Dirac representation for 
the $4 \times 4$ Dirac matrices 
\beq
\gamma^0 =\left( \begin{array}{cc} 1 & 0\\ 0 & -1 \end{array}\right), \quad
{\bf \gamma} =\left( \begin{array}{cc} 0 & {\bf \sigma}\\ -{\bf \sigma} & 0 
\end{array} \right), 
\quad{\rm and}\quad
\gamma_5 = \left( \begin{array}{cc} 0 & 1\\ 1 & 0 \end{array} \right),
\label{gamma}
\eeq
where $0$ is a $2 \times 2$ zero matrix, $1$ is a $2 \times 2$ unit matrix,
and ${\bf \sigma}$ are the three $2 \times 2$ Pauli matrices.  Let us further
write the six Plucker
coordinates (under the so-called Klein transformation)
in terms of $p_{\mu}$ with $\mu$ running over $0, 1, 2, 3$
(to be interpreted as the four-momentum
of the Dirac particle) and $m_1$ and $m_2$ (to be interpreted as two real
mass parameters) as follows:
\bea
p_{34} = + p_0 + m_1, \quad
p_{12} = - p_0 + m_1, \nonumber\\
p_{13} = + p_1 - ip_2, \quad
p_{24} = - p_1 - ip_2, \nonumber\\
p_{41} = + p_3 + m_2, \quad
p_{23} = - p_3 + m_2.
\label{pluckerp}
\eea
Then we can 
rewrite Eqs. (\ref{r1}) and (\ref{r4}) as
\beq
(\gamma^0 p_0 + {\bf \gamma} \cdot {\bf p} + m_1 + m_2 \gamma_5 ) \psi = 0,
\label{Diraceq}
\eeq
the generalized Dirac equation in energy-momentum space!  In coordinate space, 
we get
\beq
\left(i\pslash-m_1-m_2\gamma_5\right)\psi({\bf x},t)=0.
\label{e19}
\eeq

The above choice (\ref{pluckerp}) of $p_{ij}$ in terms of 
$p_{\mu}$, $m_1$ and $m_2$
is dictated by the representation of the Dirac matrices we have
adopted.  A different representation would result in a different choice.
To wit, if we use the Weyl or chiral representation for the Dirac matrices
\beq
\gamma^0 =\left( \begin{array}{cc} 0 & 1\\ 1 & 0 \end{array}\right), \quad
{\bf \gamma} =\left( \begin{array}{cc} 0 & {\bf \sigma}\\ -{\bf \sigma} & 0
\end{array} \right),              
\quad{\rm and}\quad              
\gamma_5 = \left( \begin{array}{cc} -1 & 0\\ 0 & 1 \end{array} \right),
\label{gamma1}
\eeq
we have to choose the Plucker coordinates according to 
\bea
p_{34} = + m_1 - m_2, \quad
p_{12} = + m_1 + m_2, \nonumber\\
p_{13} = + p_1 - ip_2, \quad
p_{24} = - p_1 - ip_2, \nonumber\\
p_{41} = + p_0 + p_3, \quad
p_{23} = + p_0 - p_3,
\label{pluckerp1}
\eea
to yield (\ref{Diraceq}) or (\ref{e19}).

Associated with the generalized Dirac equation (\ref{e19}) is the Hamiltonian
density for the free Dirac particle given in (\ref{e18}).  Following Bender
et al. \cite{BJR}, one can check that the Hamiltonian $H$ is not Hermitian 
because the $m_2$ term changes sign under Hermitian conjugation.  However $H$
is invariant under combined $\cal P$ and $\cal T$ reflection given by
\bea
{\cal P}\psi(\mathbf{x},t){\cal P}&=& \gamma_0\psi(-\mathbf{x},t),\cr
{\cal P}\bar\psi(\mathbf{x},t){\cal P}&=&\bar\psi(-\mathbf{x},t)\gamma_0,
\label{e31}
\eea
and
\bea
{\cal T}\psi(\mathbf{x},t){\cal T}&=&{\bf C}^{-1}\gamma_5\psi(\mathbf{x},-t),\cr
{\cal T}\bar\psi(\mathbf{x},t){\cal T}&=&\bar\psi(\mathbf{x},-t)\gamma_5{\bf C},
\label{e32}
\eea
where $\bf C$ is the charge-conjugation matrix, defined by ${\bf C}^{-1}
\gamma_\mu {\bf C}=-\gamma_\mu^{\rm T}$.  Therefore, the projective 
geometrical approach yields (at least in this particular example) 
a $\cal PT$-symmetric Hamiltonian rather than a Hermitian Hamiltonian. 
\cite{imaginarym2}

By iterating (\ref{e19}), one obtains
\beq
\left(\partial^2+\mu^2\right)\psi({\bf x},t)=0.
\label{e20}
\eeq
Thus, the physical mass that propagates under this
equation is real for $\mu^2 \geq 0$, i.e.,
\beq
m_1^2\geq m_2^2,
\label{e21}
\eeq
which defines the parametric region of unbroken $\cal PT$ symmetry. If
(\ref{e21}) is not satisfied, then the $\cal PT$ is broken. \cite{tachyons} 
And one recovers the Hermitian case only if $m_2 = 0$.  

Of course, it
would be nice if the geometrical picture alluded to in this paper could
give us some additional insight and/or predictions.  For example, one may
ask whether the values of the special cases $m_2 = 0$, which corresponds
to the standard Dirac equation, and $m_1 = m_2$, i.e., $\mu = 0$, which
marks the onset of broken $\cal PT$ symmetry, have any particular
geometrical significance. \cite{referee}  Eq. (\ref{pluckerp}) (Eq.
(\ref{pluckerp1})) shows that $m_2 = 0$ is given by the 
condition $p_{14} = p_{23}$ ($p_{12} = p_{34}$) and that $\mu = 0$ 
corresponds to $p_{12} + p_{34} = p_{41} + p_{23}$  
($p_{34} = 0$) for the Plucker 
coordinates for the case of the Dirac representation (the Weyl 
representation) of the Dirac matrices.  Unfortunately since these conditions
are representation-dependent, any potential 
geometrical significance that can be attached to these two special cases 
will probably be hard to identify.  On the other hand, as shown above, 
the projective geometrical method of ``deriving" the Dirac equation  
is very general.  
It includes
both the standard Dirac equation and the generalized Dirac equation which
yields a non-Hermitian yet $\cal PT$-symmetric Hamiltonian.  One can trace 
this feature to the simple fact that there are six Plucker 
coordinates which, in general, can {\it naturally} accommodate two types 
of masses (in addition to the four energy-momentum) for the 
spin-${\frac {1}{2}}$ particles.

Finally we note that (\ref{e20}) in the form of 
$(- p^{\mu}p_{\mu} + m_1^2 -m_2^2) \psi = 0$ is simply a reflection of the 
relation (\ref{relation1}) among the Plucker coordinates when they are
written in terms of $p^{\mu}$ , $m_1$ and $m_2$ given by either 
(\ref{pluckerp}) or (\ref{pluckerp1}).

For completeness, we should mention that Bender and collaborators \cite{BJR} 
have constructed a Hermitian Hamiltonian $h$ that corresponds to the
non-Hermitian Hamiltonian $H$ of (\ref{e18}) for  
$\mu^2 = m_1^2 - m_2^2 \geq 0$.
The two Hamiltonians are related by the
similarity transformation
\beq
h=e^{-Q/2}He^{Q/2},
\label{e6}
\eeq
where
\beq
Q=-\tanh^{-1}\!\ve\!\int\!d{\bf x}\,\psi^\dag({\bf x},t)\gamma_5\psi({\bf x},t),
\label{e33}
\eeq
with $\ve = m_2/m_1$.  The resulting $h$ is given by
\beq
h=\!\int\!d{\bf x}\,\bar\psi({\bf x},t)(-i\delslash+\mu)\psi({\bf x},t),
\label{e29}
\eeq
in agreement with (\ref{e20}). 



\begin{acknowledgments}

This work was supported in part by the U.S.~Department of Energy 
and the Bahnson Fund at the University of North Carolina.  YJN thanks 
P.~D.~Mannheim and K.~A.~Milton for useful discussions on $\cal PT$-symmetric
theories.  We thank an anonymous referee for useful suggestions.

\end{acknowledgments}


\begin{thebibliography}{10}
\bibitem{referee}
This issue is raised by an anonymous referee.
\end{thebibliography}
\end{document}